\documentclass[fleqn,usenatbib]{mnras}

\usepackage{newtxtext,newtxmath}

\usepackage[T1]{fontenc}
\usepackage{ae,aecompl}

\usepackage{graphicx}	
\def\hhe{H--He}

\title[H--He satellite in Lyman-$\alpha$ of DBA white dwarfs]{H--He collision-induced satellite in the  Lyman-$\alpha$ profile  of DBA white dwarf stars}

\author[N. F. Allard  et al.]{
Nicole F. Allard $^{1,2}$\thanks{E-mail: nicole.allard@obspm.fr},
John F. Kielkopf $^{3}$,
Siyi Xu $^{4}$,
Gr\'{e}goire Guillon $^{5}$,
\newauthor    
  Bilel Mehnen $^{6}$, 
  Roberto Linguerri $^{6}$,
  Muneerah Mogren Al Mogren $^{7}$, 
  \newauthor
  Majdi Hochlaf $^{6}$,
  Ivan Hubeny $^{8}$ 
\\ 
\hspace{1em}\\
$^{1}$GEPI, Observatoire de Paris,  Universit\'e PSL, UMR 8111, CNRS,
    61, Avenue de l'Observatoire, F-75014 Paris, France\\
$^{2}$Sorbonne Universit\'e, CNRS, UMR7095, Institut d'Astrophysique
    de Paris, 98bis Boulevard Arago, PARIS, France\\
$^{3}$Department of Physics and Astronomy, 
    University of Louisville, Louisville, Kentucky 40292 USA \\
$^{4}$Gemini Observatory, 670 N. A\'{o}hoku Place, Hilo, HI 96720 HI, USA\\
$^{5}$Laboratoire Interdisciplinaire Carnot de Bourgogne, 
    UMR6303, CNRS, Universit\'e de Bourgogne Franche Comt\'e, 
    21078 Dijon Cedex, France   \\ 
$^{6}$Universit\'{e} Gustave Eiffel, COSYS/LISIS, 5 Bd Descartes 77454, Champs sur Marne, France \\
$^{7}$Chemistry Department, Faculty of Science, King Saud University,
    PO Box 2455, Riyadh 11451, Kingdom of Saudi Arabia. \\
$^{8}$Department of Astronomy, University of Arizona, 933 N Cherry Ave, Tucson, AZ 85719  USA
}

\date{Accepted XXX. Received YYY; in original form ZZZ}

\pubyear{2019}

\begin{document}  
\label{firstpage}
\pagerange{\pageref{firstpage}--\pageref{lastpage}}

\maketitle

\begin{abstract}
The spectra of helium-dominated white dwarf stars with hydrogen in their atmosphere present a distinctive broad feature centered
around 1160~\AA\/ in the  blue wing of the  Lyman-$\alpha$ line. It is extremely
apparent in WD 1425+540 recently observed with {\it HST COS}. With
new theoretical line profiles based on ab initio atomic interaction  potentials we show that this feature is a signature of
a collision-induced satellite due to an asymptotically forbidden transition.  This quasi-molecular spectral satellite is crucial  to understanding  the
asymmetrical shape of  Lyman-$\alpha$  seen in this and other white dwarf spectra. 
Our previous work predicting this absorption feature was limited by molecular
potentials  that were not adequate to follow the atomic interactions with
spectroscopic precision  to the asymptotic limit of large separation.  A new set of
potential energy curves and   electronic dipole transition moments for the lowest
electronic states of the H--He system were developed to account  accurately for the
behaviour of the atomic interactions  at all distances, from   the chemical regime
within 1~\AA\/ out to where the radiating H atoms are not significantly perturbed
by their neighbors.
We use a  general unified theory of collision-broadened atomic
spectral lines to describe a
rigorous treatment of hydrogen Lyman-$\alpha$ with these potentials and present  a
new study of its broadening by radiative collisions of hydrogen and  neutral helium.  These results enable ab initio modeling  of radiative transport in DBA white dwarf
atmospheres.
\end{abstract} 





\section{Introduction}     \label{sec:introduction}
Theoretical studies of the effects of neutral atom collisions on atomic spectral
lines have often been hindered by our ignorance of the atomic potentials.
Even for systems as simple as H-H or H-He, the interactions and the
electric transition moments are quite
difficult to compute with the accuracy which is needed for evaluating a
complete line profile.
 The fundamental theory  of calculating the spectral line profile
\citep{allard1999}
requires  knowledge of molecular  potentials with high accuracy because
the shape and  strength of the line profile are very sensitive to the details
 of the  molecular potential curves describing the atom-atom collisions.
In  \citet{allard2009d} we made an exhaustive study of the red wing of 
Lyman-$\alpha$  line perturbed  by H--He collisions, where
we used  the potentials and electric dipole transition moments of
\citet{theodora1984} and \citet{theodora1987}.
We considered the high He densities met in cool DZ
white dwarfs and  examined the range of validity of the one-perturber
approximation widely used to calculate  the line wings.
We have shown there that the extension of the red wing of  the Lyman-$\alpha$
line seen in DZ white dwarf spectra depends strongly on the stellar temperature, while it is not 
dependent on the 
helium density. 
%
%
We also predicted a blue satellite which only very
recently has been observed in Hubble Space Telescope Cosmic Origins Spectrograph ({\it HST COS}\/) observations \citep{xu2017}.
The importance of a correct determination of the blue wing of
Lyman-$\alpha$ line to interpret   the asymmetrical shape of the    Lyman-$\alpha$ line observed with {\it COS}\/ is presented
in Sect.~\ref{sec:WD1425}.
An accurate prediction of  the satellite and consequently the full Lyman-$\alpha$ profile  requires exacting new ab initio calculations to obtain the ground
and first excited potential energy curves 
and the corresponding electric dipole transition
moments for the H--He system.
%
%
The new molecular data  in Sect.~\ref{sec:potHHe}
corroborate the prediction of a  line satellite in the Lyman-$\alpha$ profile
 \citep{allard2009d} that
 is described in Sect.~\ref{sec:ly20}.
In \citet{allard1999} we previously 
derived a classical path expression for a pressure-broadened atomic
spectral line shape that includes the effects of a radiative electric dipole
transition moment that is  dependent on the position of the  radiating atom and its dynamic neighbors.
Such a comprehensive unified  approach employing  the precise molecular data 
is fundamentally necessary  to obtain an  accurate  absorption  line profile that is valid over the full breadth of spectral line for the range of densities and temperatures found in stellar atmospheres.

\begin{figure}
\resizebox{0.46\textwidth}{!}
{\includegraphics*{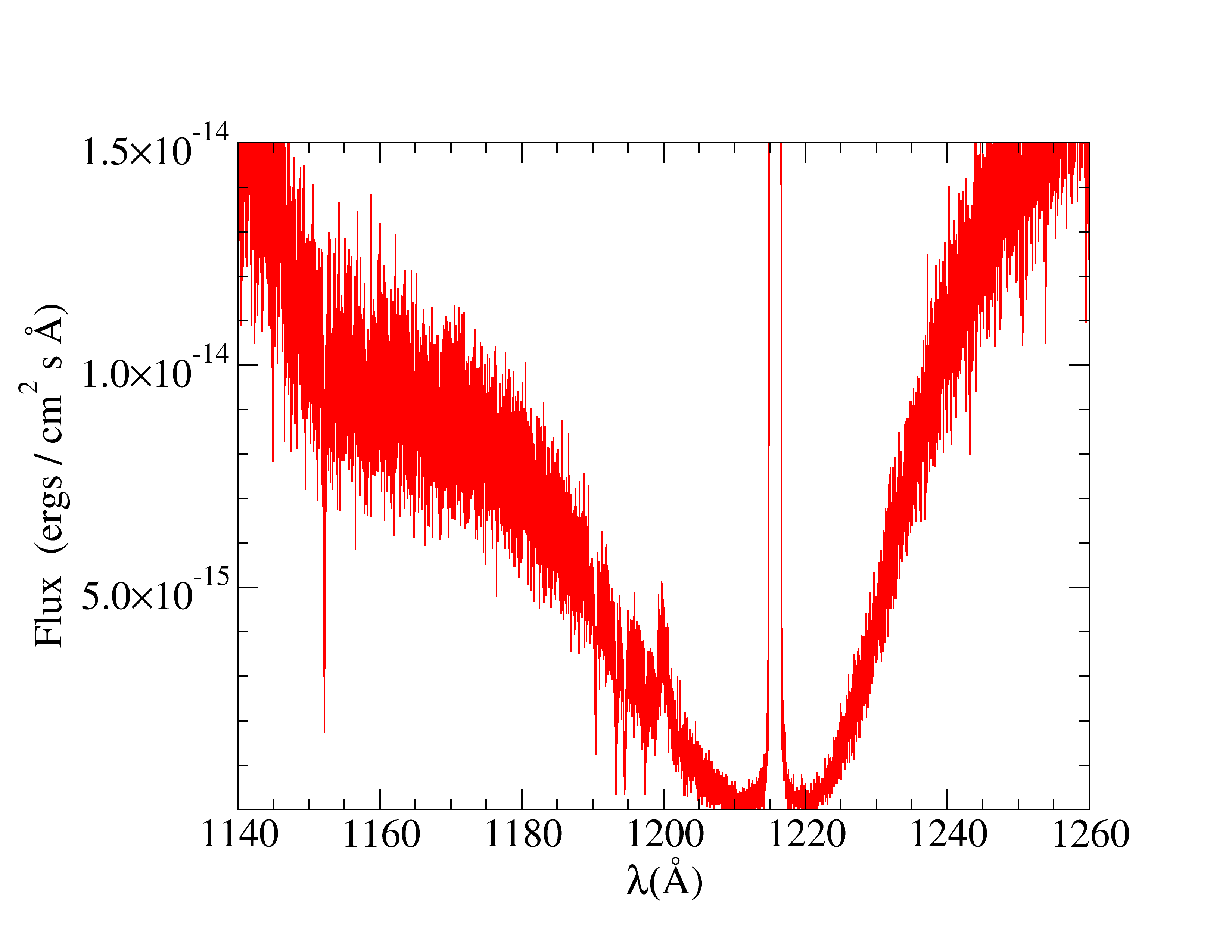}}
\caption  {{\it COS} observation of WD 1425+540.
  The broad distinctive collision-induced satellite  in the blue wing of 
 the  Lyman-$\alpha$ line about 1160~\AA\ is clearly visible\/
 \citep{xu2017}. The strong emission at the center of Lyman-$\alpha$ is from Earth's geocoronal hydrogen above the HST orbit.
 \label{fig:1425}}
\end{figure}

\section{COS observation of WD 1425+540}
\label{sec:WD1425}

WD 1425+540 (T=14,490 K, log g=7.95) is the prototype of DBA white dwarfs and
it is a helium-dominated white dwarf that also has a large amount of hydrogen
in its atmosphere \citep{bergeron2011}. It was observed with {\it HST COS}  under program 13453, and the details of observation and data reduction strategy
were reported by \citet{xu2017}. Here, we focus on the spectrum of
segment B of the G130M grating, which covers 1130-1270~\AA\/, as shown
in Fig.~\ref{fig:1425}. 
As described in \citet{xu2017}, there are two unusual features of the
Lyman-$\alpha$ profile in WD 1425+540.
First, the line profile is very asymmetric exhibiting an extend blue wing with the satellite feature as noted.  Second, previous white dwarf spectral models
cannot reproduce the strength of
Lyman-$\alpha$ and Balmer-$\alpha$ simultaneously.
The derived hydrogen abundance
is more than a factor of 10 higher from the Lyman-$\alpha$ measurement than from 
Balmer-$\alpha$. While WD 1425+540 is the most extreme case so far, these
peculiarities have been observed in other DBA white dwarfs as well, 
e.g. \citet{jura2012b}.
%
%

%
%

The asymmetry also could not be produced by white dwarf models
of \citet{xu2017} because the opacity data used for  the
Lyman-$\alpha$ profile did not take into account the quasi-molecular line
satellite predicted in \citet{allard2009d}. Once this feature is included, the observed asymmetry is reproduced  
 \citep{gansicke2018}.
The need to have both accurate data for Lyman-$\alpha$ and
for Balmer-$\alpha$ is essential  to determine  the hydrogen abundance correctly. 
The  
goal of this paper is to develop the foundation of the atomic and molecular physics needed to compute a complete profile without making ad hoc assumptions. We emphasize the importance of
accurate potentials and electric dipole transition moment data for this purpose, and here we provide that data for Lyman-$\alpha$. With new potentials of H-He we also compute a model DBA white dwarf spectrum that demonstrates their validity. 

\section{\hhe\/ diatomic potentials}
 \label{sec:potHHe}

\subsection{Methodology and benchmarks}
The lowest electronic excited states of hydrogen and helium are 
at unusually high energies for neutral atoms (> 10 eV) with respect to their ground states, and close to
the corresponding ionization thresholds.
Hydrogen with $n$= 2 or greater is a Rydberg atom in this sense
\citep{gallagher1994}.

The electronic excited states of H--He diatomic system
of interest in the present work correlate adiabatically to
those of these atoms. Therefore, for the correct description of the
electronic states of the H--He diatomic system consistent with  its isolated atomic
fragments one needs the inclusion of diffuse functions that can flexibly represent the states.
In addition to this, the computation
of the possible interactions that may occur between these electronic states
and the subsequent mixing of their wavefunctions that results in an apparent change in electric dipole transition moments, require post
Hartree-Fock multi-configurational approaches. More specifically, we used
the Complete Active Space Self Consistent Field (CASSCF)
\citep{knowles1985,werner1985}
followed by the internally contracted Multi-Reference Configuration
Interaction (MRCI)
\citep{knowles1988,werner1988,shamasundar2011}
methods as implemented in the MOLPRO 2015 package \citep{molpro2015}.
In MRCI, the complete CASSCF wave functions are used as a reference.
Furthermore, the Davidson correction (MRCI+Q) \citep{langhoff1974}
has been applied to the resulting energies to account for the lack of
size-consistency of the MRCI method. These computations were performed in
the $C_{2v}$ point group, where the $B_1$ and $B_2$ representations
were treated on equal footing.

Benchmarks on valence-Rydberg electronic states of other molecular
systems \citep{spelsberg2001,ndome2008,hochlaf2010}
showed the need to use a CASSCF active space larger than the full-valence
space. The atomic basis set for the H and He atoms had to be optimized as
well. Thus, we performed a series of benchmark computations at different
levels of accuracy to find the appropriate states for convergence.

Firstly, at the lowest level of accuracy, we adopted a
small active space of 3 electrons in 7 molecular orbitals in conjunction with
the aug-cc-pV5Z \citep{dunning1989,dunning1992} basis set.
With this approach, we found inconsistencies in the
calculated energies, especially in the asymptotic region. Indeed, with this simplest choice there is 
a large energy gap of $\sim0.45$~eV between the two equivalent dissociation limits
{\mbox H($2p \, ^2P$) + He($1s^2 \, ^1S$)}
and {\mbox H($2s \, ^2S$) + He($1s^2 \, ^1S$)}. Obviously, this gap is
unphysical since these two asymptotes should be strictly degenerate because the two H ($n=2$) states have the same energy apart from Lamb shift and negligibly small fine and hyperfine structure.
Moreover, we found a spurious second potential well ($D_e$ $\sim$ 660 cm$^{-1}$)
 in the $C \, \Sigma$ state of H--He at large internuclear separations
(for $R_{\mathrm{H-He}}$ $\sim$ 4.2~\AA\/). Thus, at this level of accuracy,
a rather poor
chemical description of the H--He molecule is obtained in spite of the relatively large size
of the MRCI computations with  $\sim$~4.3 x 10$^4$ uncontracted configuration state
functions (CSFs) per $C_{2v}$ symmetry. This may be linked to some missing
correlation energy in the MRCI wavefunctions that can be recovered by
means of larger active spaces in the reference CASSCF vector and 
by adopting more diffuse atomic basis sets.

Secondly, we tried an enlarged CASSCF active space of 3 electrons in 14
molecular orbitals  in conjunction with the aug-cc-pV6Z \citep{dunning1989,dunning1992}
basis set. In the subsequent MRCI treatment, the multi-configuration wave
functions included $\sim$~2.1 x 10$^5$ uncontracted CSFs per $C_{2v}$ symmetry.
With this ansatz, the energy difference between the above mentioned asymptotes
was reduced to  $\sim$~0.33~eV but still did not vanish. For modeling based on unified spectral line shape theory an error of this size would be unacceptable.

Finally, using the same active space as in the second series of computations,
we added a set of diffuse functions to the aug-cc-pV6Z basis set for H
and He. Hereafter, this enlarged set will be denoted as aug-cc-pV6Z$^\star$.
The exponents of the added Gaussian primitives, which were left
uncontracted, are listed in Table~\ref{tab:gaussian} in the Appendix.

This approach, compared to the
previous ones, solved all the inconsistencies mentioned above. That is, it yielded
degenerate {\mbox H($2p \, ^2P$) + He($1s^2 \, ^1S)$}
and {\mbox H($2s \, ^2S$) + He($1s^2 \, ^1S)$}
dissociation limits and no spurious potential well in the $C \,\Sigma$ state.
We note that convergence was reached at this step since a further expansion of the aug-cc-pV6Z$^\star$ set by inclusion of
more diffuse functions led to almost identical results. 
In these calculations, the MRCI wave functions
included more than $7.5 \times 10^5$ uncontracted CSFs per $C_{2v}$ symmetry species.
These  relatively large computations for such a small molecular system were necessary to obtain the precision needed to model the Lyman-$\alpha$ profile accurately.

\begin{figure}
 \centering
\resizebox{0.46\textwidth}{!}
{\includegraphics*{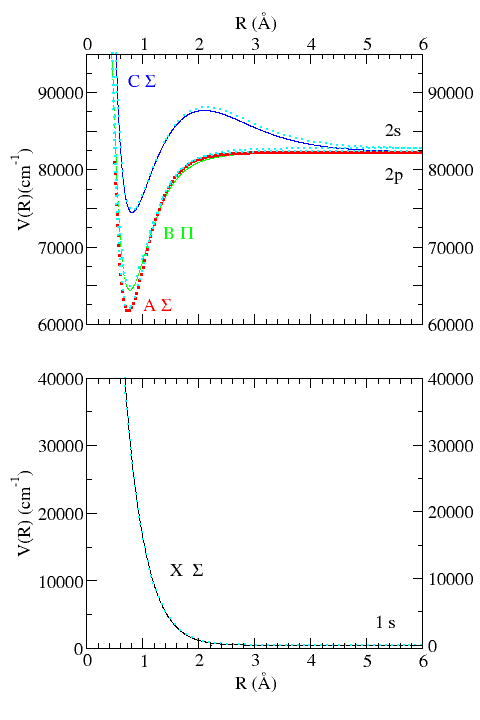}}
\caption  {Top: short-range part of the potential
  curves of the H--He molecule: $A$ (red dotted),
  $B$ (green dashed line) and $C$ (blue solid).
  Bottom: $X$ (black solid). Note the agreement at short distance with data of
  \citet{theodora1984} that are overplotted in dotted cyan.}
\label{pot}
\end{figure}

\begin{figure}
 \centering
\resizebox{0.46\textwidth}{!}
{\includegraphics*{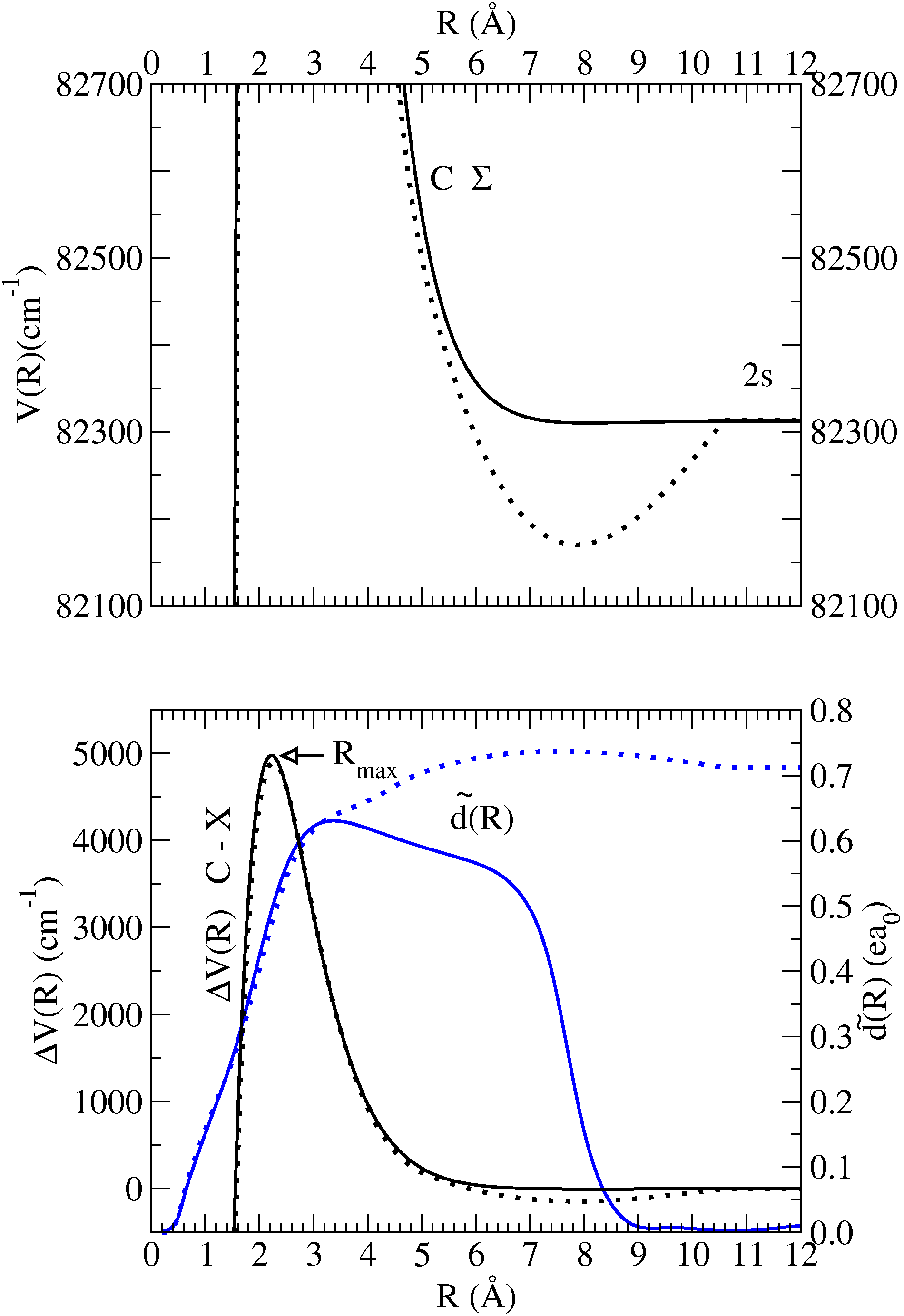}}
\caption  {Top: long range part of the $C \, \Sigma$ potential curve
    correlated with $2s$  state. This work (full line),
    \citet{theodora1984} (dotted  line).
    Bottom: $\Delta V(R)$ (black)  and $\tilde{d}(R)$ (blue) at  14500~K for the $C-X$ transition.
    The atomic separation for the maximum in the $C-X$ difference potential is $R_{\mathrm{max}}\approx 2.2\,$\AA\/ as shown in Fig.~\ref{varT}  Note that the $C-X$  transition in this work
    is forbidden asymptotically as it is a transition between the $2s$
    and $1s$ states of the free hydrogen atom at large $R$.}
\label{dip}
\end{figure}

\subsection{Potential energy curves and transition moments}
\label{sec:dipHHe}

The electronic states investigated in the present contribution correlate,
at large internuclear distances, to the
\mbox{ H($1s\;^2S$) + He($2s^2\;^1S$)},
\mbox{ H($2s\;^2P$) + He($2s^2\;^1S$)}, and
\mbox{ H($2p\;^2P$) + He($2s^2\;^1S$)}
dissociation limits
(see Fig.~\ref{pot} and Table~\ref{tab:energies} in the Appendix).
The MRCI+Q/aug-cc-pV6Z$^\star$ potential
energy curves of the four lowest electronic states of H--He, obtained with
the largest active space and basis set as described in the previous section,
are represented in Fig.~\ref{pot} as a function of the internuclear distance,
$R_{\mathrm{H-He}}$.
This figure shows that the ground state possesses a repulsive potential
correlating to the \mbox{H($1s \, ^2S$) + He($1s^2 \,^1S$)} isolated atom asymptote at large
distances. 

The  ground $X\,^2\Sigma^+$ state is repulsive at short range with a shallow well at $4\,$\AA. The excited $A\,^2\Sigma^+$, $B\,^2\Pi$ and $C\,^2\Sigma^+$
states have rather deep potential wells
in the molecular region closer than 1~\AA, and complex behavior at longer range that
can affect transition probabilities and difference potential energies in subtle ways.
We refer to these as the $X\,\Sigma$, $A\,\Sigma$, $B\,\Pi$, and $C\,\Sigma$  states, or more succinctly by the letter designations $X$, $A$, $B$, and $C$ in the following.
They correlate adiabatically to
the \mbox{H($n=2$)\, + \, He($1s^2 \,^1S$)} dissociation limits at large internuclear
separations (see Table~\ref{tab:energies} in the Appendix). 
The ordering of the assignments of labels for the states is with $A\,\Sigma$ the lowest and $C\,\Sigma$ the highest inside this close 1~\AA\/ region with wells in all the states of the order of $15\,000\;\mathrm{cm}^{-1}$ deep,
with minima  located at
$R_{\mathrm{H-He}}$ = 0.7407, 0.7686, and 0.8095~\AA\/ for the $A$, $B$ and $C$ states,
respectively (see Table~\ref{tab:constants} in the Appendix).  
While the $A$ and $B$ states have 
potentials with a simple short-range well, the $C$ state also exhibits a potential maximum of $\approx 0.666$~eV at 
$R_{\mathrm{H-He}}=2.098$~\AA.  Its  presence causes a related maximum in the $C-X$ transition difference potential energy curve which affects the blue wing of  Lyman-$\alpha$.  

Although the $C \, \Sigma$ H-He molecular state shown in Fig.~\ref{pot} is correlated asymptotically with the $2s$ atomic state, we find that at $R_{\mathrm{H-He}} < 7\;$\AA\/ the transition probability to the $X\,\Sigma$ ground state is not zero. Detailed electric dipole transition moments between the
$X \, \Sigma$ ground state and the $A \, \Sigma$, $B \,\Pi$ and $C \, \Sigma$
excited states as a function of the internuclear distance have been calculated at the
MRCI/aug-cc-pV6Z$^\star$ level.
In this calculation almost all the transition moments are rather large,
particularly for the
$C \,\Sigma$ $\leftarrow $ $A \,\Sigma$
and $B \, \Pi$ $\leftarrow $ $A \, \Sigma$ transitions,
where corresponding matrix elements of around -9.2 and -7.5 debye  (D or $10^{-18}$ statcoulomb-cm) are
calculated, respectively. Fig.~\ref{dip4} in the Appendix offers a detailed view.
These transition moments correlate to the
correct atomic values at dissociation. In particular, the
\mbox { $ \langle X \,\Sigma  | DM | C \, \Sigma \rangle$ }
matrix element of the electric dipole transition moment (DM) vanishes at large $R_{\mathrm{H-He}}$ where
the  $1s-2s$
transition in the isolated hydrogen atom is forbidden to one-photon electric dipole transitions by parity conservation.

%
%
\section{Lyman-alpha opacity}
\label{sec:ly20}
 The theory of spectral line shapes, especially the unified approach we
developed, determines the contributions of specific spectral lines to stellar opacities and  may be incorporated into stellar atmosphere  models to make accurate synthesis of stellar spectra possible.
The line shape theory accounts for neutral atom broadening and shift in  both
the centers of spectral lines and their extreme wings with one consistent
treatment without ad hoc assumptions about the line shape or potentials. Complete details and the derivation of the theory are provided by~\citet{allard1999}.
The spectrum,  $I(\Delta \omega)$, is
the Fourier transform (FT) of a electric dipole transition  autocorrelation
function, $\Phi (s)$. 
For a perturber density $n_p$, we have
\begin{equation}
\Phi(s) = e^{-n_{p}g(s)} \; ,
\label{eq:intg}
\end{equation}
where the decay of the autocorrelation function with time leads to atomic line
broadening. (See Eq.~(121) of  \citet{allard1999}.)
Our  approach introduces the concept of a modulated electric dipole transition
moment $\tilde{d}_{if}(R(t))$ into the line shape calculation. 
\begin{equation}
\tilde{d}_{if}[R(t)] = d_{if}[R(t)]e^{-\frac {V_i [R(t)]}{2kT}} \; \; ,
 \label{eq:dip}
\end{equation}
where the potential energy for the initial state is
\begin{equation}
V_{i}(R) = E_i(R)-E_i^{\infty} \; \; .
\end{equation}
The difference potential energy $\Delta V (R)$ for a transition
$if$ is
\begin{equation}
\Delta V(R)=V_{if}(R) = V_{f }(R) - V_{i}(R) \; \;  .
 \label{eq:deltaV}
\end{equation}
The Boltzmann factor $e^{-\frac {V_i (R)}{2kT}}$ in Eq.~(\ref{eq:dip})
appears because the
perturbing atoms or ions are in thermal equilibrium with the radiating
atom which affects the probability of finding them initially at a given $R$.
This  treatment results in
Lyman series line wing profiles that exhibit a sensitive dependence on temperature.
We had to use electric dipole moments modulated by the Boltzmann factor in
the comparison of emission spectra of Lyman-$\alpha$ \citep{kielkopf1998}
 and Balmer~$\alpha$ \citep{kielkopf2002} measured in laboratory.

\subsection{Study of the characteristics of  the line satellite }
\label{sec:CI}

In \citet{allard2009d} we predicted a  line satellite at 1157~\AA\/ in
spectra computed for the temperature range of cool DZ white dwarfs with potentials published in ~\citet{theodora1984}.  
However, we noticed an unexpected  well of about 150~cm$^{-1}$
(upper Fig.~\ref{dip})
in the potential energy of the $C \,\Sigma$ state at $R \sim 8$~\AA\/ which may be related to the choice of basis states and has no clear physical origin.  In this work we use the new ab initio calculations of the  potentials over the full range of distances $R$ between the H and He atoms since convergence at large $R$ is now  reached.  The long range well of the $C \, \Sigma $ state  of \citet{theodora1984} and \citet{theodora1987} potentials is not
found in these new calculations as we see in Fig.~\ref{dip}.

  \begin{figure}
\centering
\vspace{8mm}
\includegraphics[width=8cm]{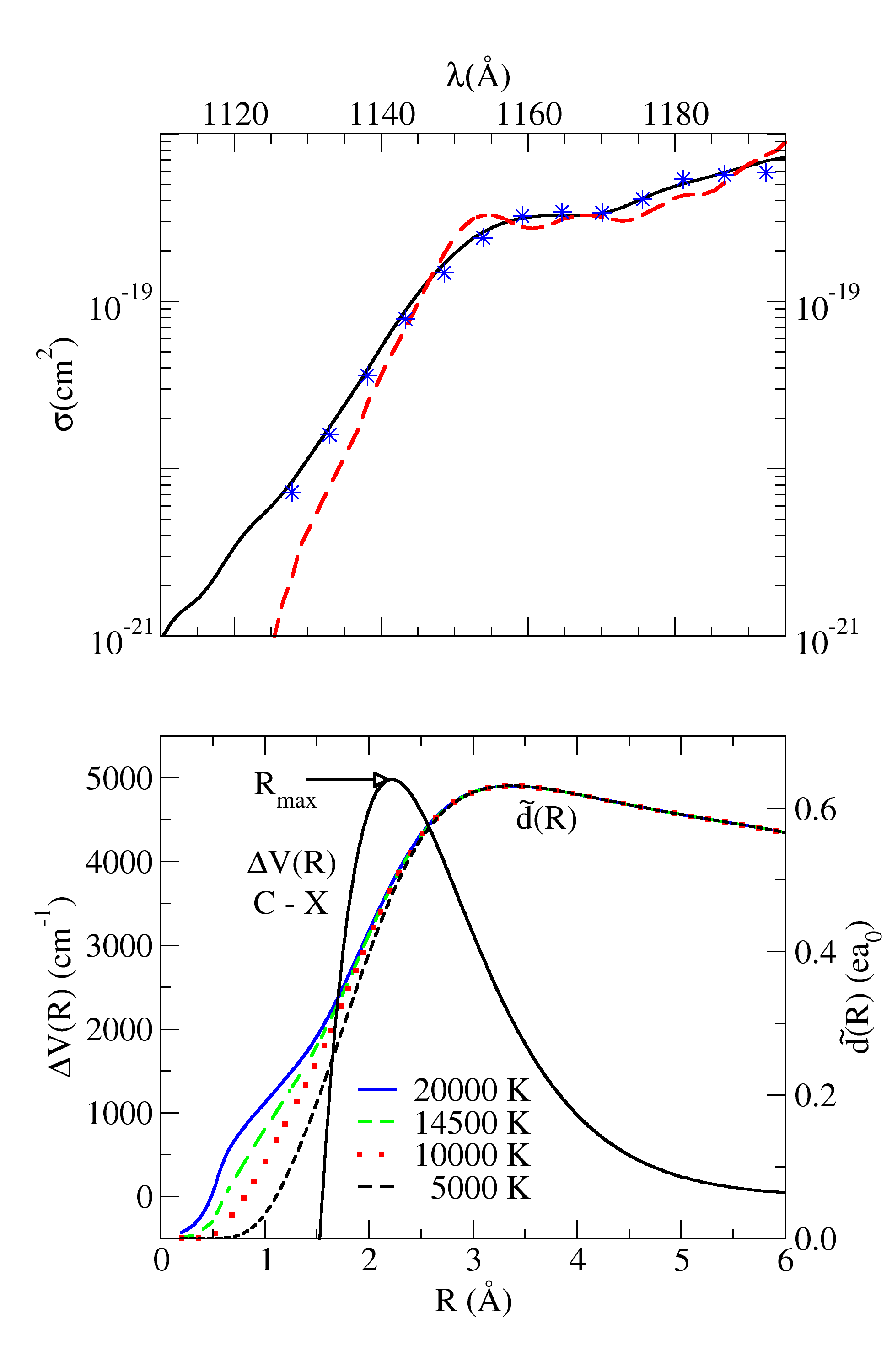}
\caption{Top: variation with temperature of the  line satellite.
  The  He density is $1 \times 10^{20}$  cm$^{-3}$, the  temperatures are
  14500~K (full black line), $20\,000$~K (blue stars), and
  $5\,000$~K (red dashed line).
  Bottom: for the $C-X$ transition, $\Delta V (R)$ (black solid) and $\tilde{d(R)}$ at 5000~K (black solid), $10\,000$~K (red dotted), 14500~K (green dashed), and 20000~K (blue solid). At the highest temperatures
  the He can reach the inner regions of the lower state $X\,^2\Sigma$ potential and enhance the transition probability.}
\label{varT}
  \end{figure}

The prediction  of a satellite in the blue wing of
the  H--He line profile is related to a potential maximum at $R=2.1$~\AA\/
(see  Sect.~\ref{sec:dipHHe})
of the $C \, \Sigma$ state. This leads to a maximum
of the potential energy difference $\Delta V(R)$ in Eq.~(\ref{eq:deltaV})
for this transition shown in Fig.~\ref{dip}.

 The  unified theory  predicts that line satellites
 will be centered
 periodically at frequencies corresponding to integer multiples of
 the extrema of  $\Delta V(R)$. In the quasi-static limit the first satellite on the line  would be  
 at $\Delta \omega = 5\,000$~cm$^{-1}$ corresponding 
 to $\lambda \sim 1150$~\AA\/ on the blue side of Lyman-$\alpha$.
 In this case the maximum in $\Delta V$ occurs at rather small
internuclear distance, and is quite sharp. The correspondingly short
duration of the close collision
leads to a broad satellite centered at
$\lambda$ $\sim$ 1160~\AA\/ for T=$14\,500$~K (Fig.~\ref{varT}).

\subsection{Temperature and density dependence}
\label{lyhe}

For a lower temperature, $T=5\,000$~K (Fig.~\ref{varT}),  the duration of the
collision is longer,
and the line satellite at $\lambda \sim 1153$~\AA\/ is sharper and
closer to the predicted quasi-static position than at higher temperatures.
The oscillations which appear on the red side of the quasi-molecular
satellite are due to interference effects 
described by \citet{royer1971b} and \citet{sando1973}.  They
depend on the relative velocity and therefore on temperature. Consequently velocity averaging would moderate their amplitude in observed spectra.
At temperatures below $10\,000$~K  the blue wing of Lyman-$\alpha$ shortward of
1150~\AA\/ becomes significantly more transparent than at higher temperature,  
an order of magnitude effect below 1120~\AA. Thus this far blue wing is a sensitive indicator of temperature in cool helium-rich  WD atmospheres.

\begin{figure}
\centering
\vspace{8mm}
\bigskip
\includegraphics[width=8cm]{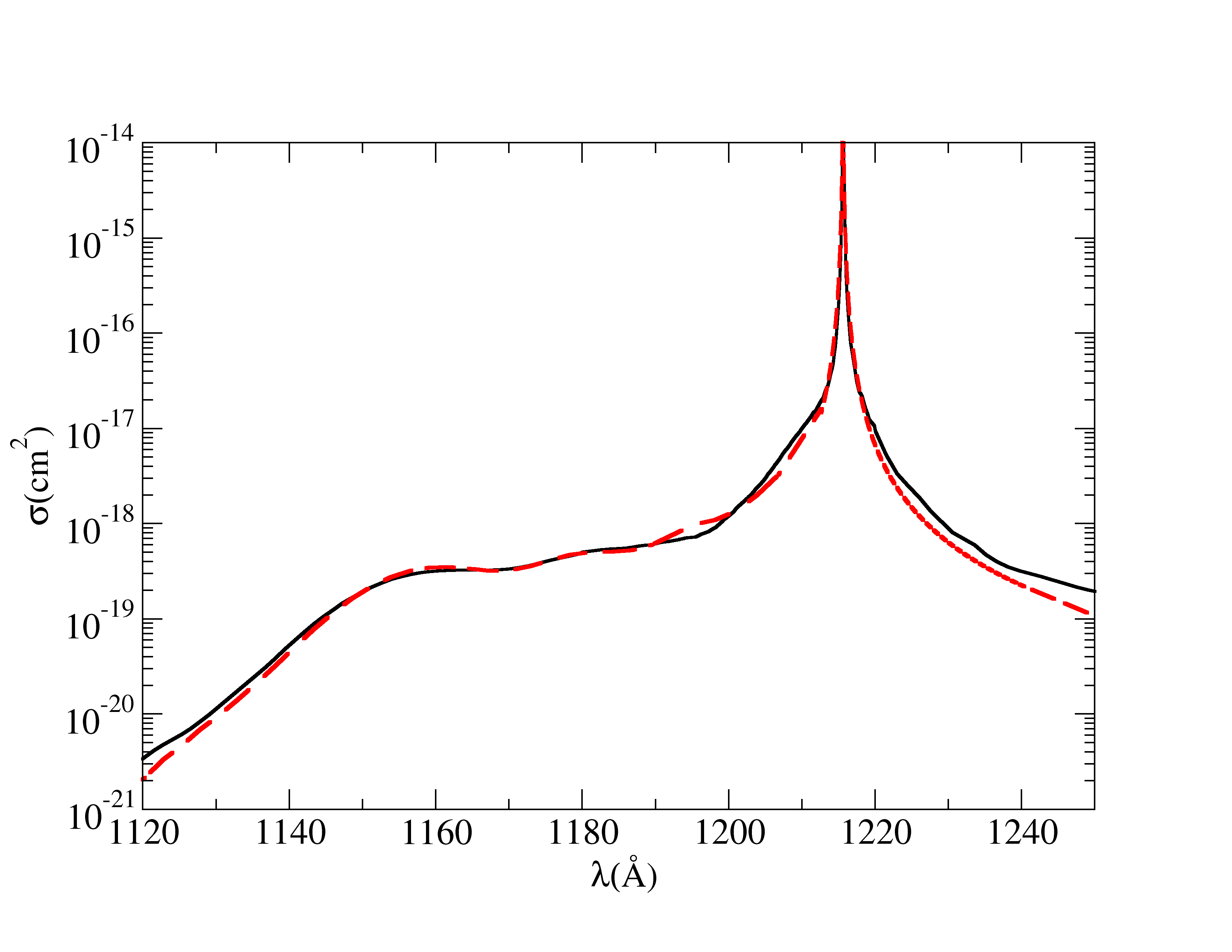}
\caption{Comparison  of the  unified line  profile using the dipole moments
  of this work  (black  line)
  with the line  profile using dipole moments of ~\citet{theodora1987} (red dashed  line) .
  The He density  is $10^{20}$  cm$^{-3}$ and
  the  temperature is 14500~K. }
\label{lyman20}
\end{figure}

 The satellite amplitude depends on the value of the electric dipole transition moment through the  region of the potential extremum responsible for the
satellite and on the position of this extremum.
The blue  line wings shown in Fig.~\ref{varT}  are  unchanged in the range
$14\,500$ to $20\,000$~K as there is no  change with $T$
of $\tilde{d}_{if}[R(t)]$ in the
internuclear distance where the
potential difference  goes through a maximum. $\tilde{d}_{if}[R(t)]$ at $14\,500$~K 
for  the $C-X$ transition is  also plotted in Fig.~\ref{dip}. 
 In the former work we used  electric dipole transition moments
 of ~\citet{theodora1987} where the $C-X$ transition was allowed.
 Nevertheless the amplitude and position of the line satellite are
 unchanged as  they are due to a range of internuclear distance where the
 potentials and the dipole moments are almost identical as we see in Fig.~\ref{lyman20}.
 The main difference between the two potentials concerns the red wing which is lowered using dipole moments of ~\citet{theodora1987} where the $A-X$ transition was forbidden.

In summary the unified line profile calculation leads to a flat 
blue wing due to a line satellite.
The resulting asymmetry of the Lyman-$\alpha$ line can be easily appreciated
in Fig.~\ref{lyman20} the blue side of the line is wider than the red side.
Measured at the strength 
of the broad collision-induced 1160~\AA\/ satellite, the asymmetry ratio
of the width on the blue side to that on the red
is as large as 2.2.  
Consequently, the near wing is clearly far different from a symmetric
Lorentzian 
because the satellite is rather close to the isolated atom line center. This 
was also the case for the Mg~b triplet perturbed by He \citep{allard2016a}.
The existence of the asymmetrical shape of these line profiles depends
 strongly on the  maximum value 
  of the potential energy difference $\Delta V(R)$ 
  which predicts the position of the line satellite and on the atomic collision energies at the temperatures of interest. 
These results enable computing atmosphere models and synthetic spectra which we compare to an HST COS observation of WD 1425+540 
in Section~\ref{sec:model}.

\section{Model atmosphere and synthetic white dwarf spectrum }
\label{sec:model} 

%
%
%
To demonstrate the importance of a proper treatment of He perturbers on
hydrogen lines,
synthetic spectra of the white dwarf WD 1425+540 were computed using the
stellar atmosphere code TLUSTY (version 207) for computing  the
atmospheric structure, and a companion program SYNSPEC (version 53)
for generating detailed synthetic spectra. For a description of the
previous versions (205 and 51) see the works of 
\citet{hubeny2017} and \cite{hubeny2011a,hubeny2011b}.
This procedure allows us to study the effect of the H/He ratio on the spectrum, and the development of line wings, 
though it is not fully self-consistent with the stellar atmosphere model since that would
require a treatment of He I optical lines as well.
We have computed a number of H-He models, with
the basic model parameters, $T_{\rm eff}=14,410$ K and $\log g=7.89$,
from 
\citet{gansicke2018},
and with varying He/H ratio. For
treating the electron and proton broadening of the hydrogen lines we
used 
\citet{tremblay2009}
data. 
 The He/H ratio was adjusted to obtain a reasonable agreement by eye with the observed spectrum, and we found a nominal ratio of $4\times10^3$ ($\log(N_\mathrm{H}/N_\mathrm{He}) \approx -3.6$) fitted the observed profile well.   \citet{liebert1979} found 3.7 from a ground-based H$\beta$ profile, and recently  \citet{gansicke2018} analyzed the L$\alpha$ profile and adopted a somewhat larger
 $\log(N_\mathrm{H}/N_\mathrm{He}) \approx -4.0\pm0.20$. 
 
 
 
\begin{figure}
\centering
\vspace{8mm}
\bigskip
\includegraphics[width=8cm]{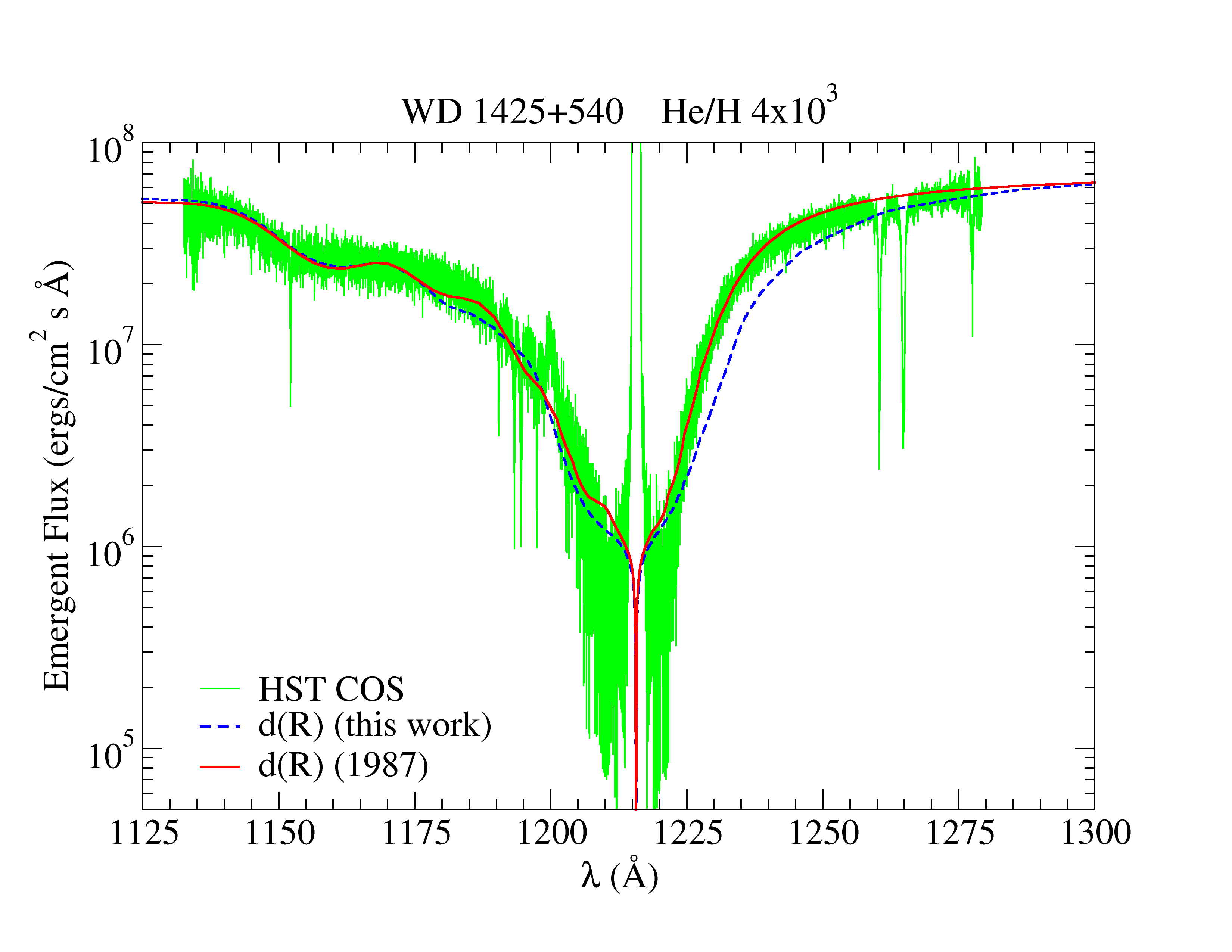}
\caption{The observed spectrum of WD 1425+540 (also see Fig.~\ref{fig:1425}) compared with a synthetic white dwarf spectrum in the Lyman-$\alpha$ region.  The synthetic spectrum is computed with TLUSTY and SYNSPEC for a temperature of 14\,500~K and a He/H ratio of $4\times10^3$ using the unified line  profile with the potentials of
this work. For the 
dipole moments of \citet{theodora1987} (red solid line)  the $A-X$ transition is forbidden and its contributions to the opacity are suppressed.  For the dipole moments of this work (blue dashed line),  the $A-X$ transition contributes in the red wing of the model but is absent in the observed spectrum.
  \label{fig:cos_mh_comp}}
\end{figure}

 The potential energies for the $n=1$ and $n=2$ electronic states H-He  that were used in our models are the ones described in this paper. Stellar opacities were computing using H-He  electric  dipole moments from the previous work of  \citet{theodora1987} in which  the $A-X$ transition is forbidden,  and also using new dipole transition moments from this work in which  the $A-X$ transition is allowed.   As shown in Fig.~\ref{fig:cos_mh_comp}, the observed red wing of Lyman-$\alpha$ is consistent with a suppressed $A-X$ transition probability  in the region of atomic separation with difference potential energy that would contribute.
 
 We conclude that the additional basis states used for the new ab initio potentials improve the calculation of the potential energy curves, but may not capture the dipole transition moments of the real H-He system correctly for the $A-X$ transition.  However the combination of  this work's potentials and the dipole moments of \citet{theodora1987} achieve a remarkable fit in Fig.~\ref{fig:cos_mh_comp} to the HST COS spectrum of WD 1425+540 when incorporated into the unified  line shape theory we described here. 

%
%

\section{Conclusions}

%
%

The Lyman-$\alpha$ region of the spectrum of a helium-rich white dwarf with hydrogen in its atmosphere is determined by the changes in  transition energy and transition probability during the H-He collisions that broaden the atomic spectral line.
We developed new H-He potential energies and  transition dipole moments for the hydrogen $1s$, $2s$, and $2p$ states as input data for a unified theory calculation of the profile of WD 1425+540 to test the potentials and dipole moments, and to confirm the origin of the short-wavelength ``blue'' satellite. We found that the spectral line profile from the new molecular data has a satellite feature in the blue wing that agrees with previous work. These results  provide a benchmark implementation of ab initio atomic and molecular potentials for the most basic neutral non-resonant atom-atom pair relevant to  stellar atmosphere models.  The new calculations show how the profile depends on  the variation of the electric dipole transition moment and interaction potential energy with  atomic separation.  A comparison with the observed spectrum of WD 1425+540 was made by using these theoretical opacities in a stellar atmosphere and spectrum synthesis code.  While it was not our goal to refine the stellar model based on the new theoretical data, the profiles reproduce the observed spectrum 
with a reasonable He/H ratio.  Further, the absence of an extended red wing of Lyman-$\alpha$ in the observed spectrum suggests that the states of the difference potential that could contribute to that region have the reduced transition dipole moment that was found in previous molecular models.  The new work presented here shows clearly that there is an opportunity to use stellar spectra to improve the atomic and molecular physics, ultimately to yield better models for astrophysical applications. For H--He, the $A-X$ transition dipole moment remains uncertain.

The blue wing of Lyman-$\alpha$ is sensitive to He density and the structure and temperature of the stellar atmosphere, with a profile that for wavelengths shortward of $1150\,$\AA\/ will have reduced opacity from regions with temperatures under $10\,000\,$K.  
Profiles computed with a  
unified theory of collision broadening based on accurate data from ab initio molecular physics take  into account the strong dependence of  the amplitude of the electric dipole transition moment on atom-atom separation ($R$)
where the potential energy change $\Delta V (R)$ is an extremum. Incorporated into model atmospheres, this dependence may be used to probe white dwarf or stellar atmospheres for density and temperature.  
This emphasizes the importance
of  the accuracy of both the potential energies  and the electric dipole
transition moments
for the line shape calculations that have traditionally assumed electric dipole transition moments are constant \citep{allard1982,allard1992,allard1994}.

The effect of collision broadening is
central to understanding  the opacity of stellar atmospheres, yet there have been only a few definitive comparisons with experimental work for atomic H.
\citep{kielkopf1995,kielkopf1998,kielkopf2004}.
This is because of the difficulty of creating an environment
 in a laboratory experiment simulating a stellar atmosphere with accurate
 diagnostics.
On the theoretical side, the maturing capability of ab initio  methods 
now offers the possibility of accurately computing the interaction of
H with H 
\citep{drira1999,spielfiedel2003,spielfiedel2004} and H with He atoms (this work).
While an  accurate determination of the broadening 
of Balmer~$\alpha$ with high density atomic hydrogen (that is H--H) has been done by \citet{allard2008a},
nothing comparable exists for H--He.
Our calculations reported in \citet{allard2008a} support the results
 of \citet{barklem2000,barklem2002}
that the \citet{ali1966} theory underestimates the actual line width.
Recent laboratory measurements show a similar result at high density
in environments comparable to white dwarf atmospheres \citep{kielkopf2014}.
It would be possible now to  similarly improve the calculation of 
 Balmer-$\alpha$ broadening and its contribution to the  full white dwarf
 opacity model. 
A major improvement to comprehensive theoretical models for DBA white dwarf
spectra is within reach that would  determine H-He molecular data for
$n=3$ excited states, and use 
those to compute accurate Balmer-$\alpha$ profiles
under white dwarf atmosphere conditions. Such results would help understanding  the differences in stellar parameters that are found from Balmer and Lyman line profiles.
 In conclusion, complete unified line profiles based on accurate
 atomic and molecular physics for both the Lyman-$\alpha$ and Balmer-$\alpha$
 lines should be incorporated into the analysis of DBA white dwarf spectra
 to derive  the  hydrogen abundance.

\section*{acknowledgements}
The paper was based on observations made with the NASA/ESA Hubble Space
Telescope under program 13453, obtained from the data archive at the Space
Telescope Science Institute. STScI is operated by the Association of
Universities for Research in Astronomy, Inc. under NASA contract NAS 5-26555.
We thank the COST Action CM1405 MOLecules in Motion (MOLIM) of the European
Community for support.
The authors would like to extend their sincere appreciation to the Deanship of Scientific Research at King Saud University for funding the research through the Research Group Project No. RGP-333.
This work was supported by the CNRS program 
Physique et Chimie du Milieu Interstellaire (PCMI) co-funded by
the Centre National d'Etudes Spatiales (CNES).




\bibliographystyle{mnras}   
\bibliography{lyman}    

\begin{thebibliography}{}
\makeatletter
\relax
\def\mn@urlcharsother{\let\do\@makeother \do\$\do\&\do\#\do\^\do\_\do\%\do\~}
\def\mn@doi{\begingroup\mn@urlcharsother \@ifnextchar [ {\mn@doi@}
  {\mn@doi@[]}}
\def\mn@doi@[#1]#2{\def\@tempa{#1}\ifx\@tempa\@empty \href
  {http://dx.doi.org/#2} {doi:#2}\else \href {http://dx.doi.org/#2} {#1}\fi
  \endgroup}
\def\mn@eprint#1#2{\mn@eprint@#1:#2::\@nil}
\def\mn@eprint@arXiv#1{\href {http://arxiv.org/abs/#1} {{\tt arXiv:#1}}}
\def\mn@eprint@dblp#1{\href {http://dblp.uni-trier.de/rec/bibtex/#1.xml}
  {dblp:#1}}
\def\mn@eprint@#1:#2:#3:#4\@nil{\def\@tempa {#1}\def\@tempb {#2}\def\@tempc
  {#3}\ifx \@tempc \@empty \let \@tempc \@tempb \let \@tempb \@tempa \fi \ifx
  \@tempb \@empty \def\@tempb {arXiv}\fi \@ifundefined
  {mn@eprint@\@tempb}{\@tempb:\@tempc}{\expandafter \expandafter \csname
  mn@eprint@\@tempb\endcsname \expandafter{\@tempc}}}

\bibitem[\protect\citeauthoryear{{Ali} \& {Griem}}{{Ali} \&
  {Griem}}{1966}]{ali1966}
{Ali} A.~W.,  {Griem} H.~R.,  1966, \mn@doi [Physical Review]
  {10.1103/PhysRev.144.366}, \href
  {http://adsabs.harvard.edu/abs/1966PhRv..144..366A} {144, 366}

\bibitem[\protect\citeauthoryear{{Allard} \& {Christova}}{{Allard} \&
  {Christova}}{2009}]{allard2009d}
{Allard} N.~F.,  {Christova} M.,  2009, \mn@doi [New Astron. Rev.]
  {10.1016/j.newar.2009.07.007}, \href
  {http://adsabs.harvard.edu/abs/2009NewAR..53..252A} {53, 252}

\bibitem[\protect\citeauthoryear{Allard \& Kielkopf}{Allard \&
  Kielkopf}{1982}]{allard1982}
Allard N.~F.,  Kielkopf J.~F.,  1982, Rev. Mod. Phys., 54, 1103

\bibitem[\protect\citeauthoryear{Allard \& Koester}{Allard \&
  Koester}{1992}]{allard1992}
Allard N.~F.,  Koester D.,  1992, A\&A, 258, 464

\bibitem[\protect\citeauthoryear{Allard, Koester, Feautrier  \&
  Spielfiedel}{Allard et~al.}{1994}]{allard1994}
Allard N.~F.,  Koester D.,  Feautrier N.,   Spielfiedel A.,  1994, A\&A Suppl.,
  108, 417

\bibitem[\protect\citeauthoryear{Allard, Royer, Kielkopf  \& Feautrier}{Allard
  et~al.}{1999}]{allard1999}
Allard N.~F.,  Royer A.,  Kielkopf J.~F.,   Feautrier N.,  1999, Phys. Rev. A,
  60, 1021

\bibitem[\protect\citeauthoryear{Allard, Kielkopf, Cayrel  \& van~'t
  Veer-Menneret}{Allard et~al.}{2008}]{allard2008a}
Allard N.~F.,  Kielkopf J.~F.,  Cayrel R.,   van~'t Veer-Menneret C.,  2008,
  A\&A, 480, 581

\bibitem[\protect\citeauthoryear{Allard, Leininger, Gad\'ea, Brousseau-Couture
  \& Dufour}{Allard et~al.}{2016}]{allard2016a}
Allard N.~F.,  Leininger T.,  Gad\'ea F.~X.,  Brousseau-Couture V.,   Dufour
  P.,  2016, \mn@doi [A\&A] {10.1051/0004-6361/201527826}, 588, A142

\bibitem[\protect\citeauthoryear{{Barklem}, {Piskunov}  \& {O'Mara}}{{Barklem}
  et~al.}{2000}]{barklem2000}
{Barklem} P.~S.,  {Piskunov} N.,   {O'Mara} B.~J.,  2000, \aap, \href
  {http://adsabs.harvard.edu/abs/2000A%26A...363.1091B} {363, 1091}

\bibitem[\protect\citeauthoryear{{Barklem}, {Stempels}, {Allende Prieto},
  {Kochukhov}, {Piskunov}  \& {O'Mara}}{{Barklem} et~al.}{2002}]{barklem2002}
{Barklem} P.~S.,  {Stempels} H.~C.,  {Allende Prieto} C.,  {Kochukhov} O.~P.,
  {Piskunov} N.,   {O'Mara} B.~J.,  2002, \mn@doi [\aap]
  {10.1051/0004-6361:20020163}, \href
  {http://adsabs.harvard.edu/abs/2002A%26A...385..951B} {385, 951}

\bibitem[\protect\citeauthoryear{{Bergeron} et~al.,}{{Bergeron}
  et~al.}{2011}]{bergeron2011}
{Bergeron} P.,  et~al., 2011, \mn@doi [\apj] {10.1088/0004-637X/737/1/28},
  \href {http://adsabs.harvard.edu/abs/2011ApJ...737...28B} {737, 28}

\bibitem[\protect\citeauthoryear{{Drira}}{{Drira}}{1999}]{drira1999}
{Drira} I.,  1999, \mn@doi [Journal of Molecular Spectroscopy]
  {10.1006/jmsp.1999.7934}, \href
  {http://adsabs.harvard.edu/abs/1999JMoSp.198...52D} {198, 52}

\bibitem[\protect\citeauthoryear{{Dunning}}{{Dunning}}{1989}]{dunning1989}
{Dunning} Jr. T.~H.,  1989, \mn@doi [\jcp] {10.1063/1.456153}, \href
  {http://adsabs.harvard.edu/abs/1989JChPh..90.1007D} {90, 1007}

\bibitem[\protect\citeauthoryear{{Gallagher}}{{Gallagher}}{1994}]{gallagher1994}
{Gallagher} T.~F.,  1994, {Rydberg Atoms}.
Cambridge University Press, Cambridge, U.K.

\bibitem[\protect\citeauthoryear{{G{\"a}nsicke}, {Koester}, {Farihi}  \&
  {Toloza}}{{G{\"a}nsicke} et~al.}{2018}]{gansicke2018}
{G{\"a}nsicke} B.~T.,  {Koester} D.,  {Farihi} J.,   {Toloza} O.,  2018,
  \mn@doi [\mnras] {10.1093/mnras/sty2526}, \href
  {http://adsabs.harvard.edu/abs/2018MNRAS.481.4323G} {481, 4323}

\bibitem[\protect\citeauthoryear{{Hochlaf}, {Ndome}, {Hammout{\`e}ne}  \&
  {Vervloet}}{{Hochlaf} et~al.}{2010}]{hochlaf2010}
{Hochlaf} M.,  {Ndome} H.,  {Hammout{\`e}ne} D.,   {Vervloet} M.,  2010,
  \mn@doi [Journal of Physics B Atomic Molecular Physics]
  {10.1088/0953-4075/43/24/245101}, \href
  {http://adsabs.harvard.edu/abs/2010JPhB...43x5101H} {43, 245101}

\bibitem[\protect\citeauthoryear{{Hubeny} \& {Lanz}}{{Hubeny} \&
  {Lanz}}{2011a}]{hubeny2011a}
{Hubeny} I.,  {Lanz} T.,  2011a, {TLUSTY: Stellar Atmospheres, Accretion Disks,
  and Spectroscopic Diagnostics} (\mn@eprint {ascl} {1109.021})

\bibitem[\protect\citeauthoryear{{Hubeny} \& {Lanz}}{{Hubeny} \&
  {Lanz}}{2011b}]{hubeny2011b}
{Hubeny} I.,  {Lanz} T.,  2011b, {Synspec: General Spectrum Synthesis Program}
  (\mn@eprint {ascl} {1109.022})

\bibitem[\protect\citeauthoryear{Hubeny \& Lanz}{Hubeny \&
  Lanz}{2017}]{hubeny2017}
Hubeny I.,  Lanz T.,  2017, A brief introductory guide to TLUSTY and SYNSPEC
  (\mn@eprint {arXiv} {1706.01859})

\bibitem[\protect\citeauthoryear{{Jura}, {Xu}, {Klein}, {Koester}  \&
  {Zuckerman}}{{Jura} et~al.}{2012}]{jura2012b}
{Jura} M.,  {Xu} S.,  {Klein} B.,  {Koester} D.,   {Zuckerman} B.,  2012,
  \mn@doi [\apj] {10.1088/0004-637X/750/1/69}, \href
  {http://adsabs.harvard.edu/abs/2012ApJ...750...69J} {750, 69}

\bibitem[\protect\citeauthoryear{{Kendall}, {Dunning}  \& {Harrison}}{{Kendall}
  et~al.}{1992}]{dunning1992}
{Kendall} R.~A.,  {Dunning} Jr. T.~H.,   {Harrison} R.~J.,  1992, \mn@doi
  [\jcp] {10.1063/1.462569}, \href
  {http://adsabs.harvard.edu/abs/1992JChPh..96.6796K} {96, 6796}

\bibitem[\protect\citeauthoryear{Kielkopf \& Allard}{Kielkopf \&
  Allard}{1995}]{kielkopf1995}
Kielkopf J.~F.,  Allard N.~F.,  1995, ApJ, 450, L75

\bibitem[\protect\citeauthoryear{Kielkopf \& Allard}{Kielkopf \&
  Allard}{1998}]{kielkopf1998}
Kielkopf J.~F.,  Allard N.~F.,  1998, Phys. Rev. A, 58, 4416

\bibitem[\protect\citeauthoryear{{Kielkopf} \& {Allard}}{{Kielkopf} \&
  {Allard}}{2014}]{kielkopf2014}
{Kielkopf} J.~F.,  {Allard} N.~F.,  2014, \mn@doi [Journal of Physics B Atomic
  Molecular Physics] {10.1088/0953-4075/47/15/155701}, \href
  {http://adsabs.harvard.edu/abs/2014JPhB...47o5701K} {47, 155701}

\bibitem[\protect\citeauthoryear{{Kielkopf}, {Allard}  \&
  {Decrette}}{{Kielkopf} et~al.}{2002}]{kielkopf2002}
{Kielkopf} J.~F.,  {Allard} N.~F.,   {Decrette} A.,  2002, \mn@doi [European
  Physical Journal D] {10.1140/e10053-002-0006-7}, \href
  {http://adsabs.harvard.edu/abs/2002EPJD...18...51K} {18, 51}

\bibitem[\protect\citeauthoryear{{Kielkopf}, {Allard}  \& {Huber}}{{Kielkopf}
  et~al.}{2004}]{kielkopf2004}
{Kielkopf} J.~F.,  {Allard} N.~F.,   {Huber} J.,  2004, \mn@doi [\apjl]
  {10.1086/423895}, \href {http://adsabs.harvard.edu/abs/2004ApJ...611L.129K}
  {611, L129}

\bibitem[\protect\citeauthoryear{{Knowles} \& {Werner}}{{Knowles} \&
  {Werner}}{1985}]{knowles1985}
{Knowles} P.~J.,  {Werner} H.-J.,  1985, \mn@doi [Chemical Physics Letters]
  {10.1016/0009-2614(85)80025-7}, \href
  {http://adsabs.harvard.edu/abs/1985CPL...115..259K} {115, 259}

\bibitem[\protect\citeauthoryear{{Knowles} \& {Werner}}{{Knowles} \&
  {Werner}}{1988}]{knowles1988}
{Knowles} P.~J.,  {Werner} H.-J.,  1988, \mn@doi [Chemical Physics Letters]
  {10.1016/0009-2614(88)87412-8}, \href
  {http://adsabs.harvard.edu/abs/1988CPL...145..514K} {145, 514}

\bibitem[\protect\citeauthoryear{{Kramida}}{{Kramida}}{2010}]{kramida2010}
{Kramida} A.~E.,  2010, \mn@doi [Atomic Data and Nuclear Data Tables]
  {10.1016/j.adt.2010.05.001}, \href
  {http://adsabs.harvard.edu/abs/2010ADNDT..96..586K} {96, 586}

\bibitem[\protect\citeauthoryear{Langhoff \& Davidson}{Langhoff \&
  Davidson}{1974}]{langhoff1974}
Langhoff S.~R.,  Davidson E.~R.,  1974, J. Quant. Chem., 8, 61

\bibitem[\protect\citeauthoryear{{Liebert}, {Gresham}, {Hege}  \&
  {Strittmatter}}{{Liebert} et~al.}{1979}]{liebert1979}
{Liebert} J.,  {Gresham} M.,  {Hege} E.~K.,   {Strittmatter} P.~A.,  1979,
  \mn@doi [Astronomical Journal] {10.1086/112584}, 84, 1612

\bibitem[\protect\citeauthoryear{{Ndome}, {Hochlaf}, {Lewis}, {Heays}, {Gibson}
   \& {Lefebvre-Brion}}{{Ndome} et~al.}{2008}]{ndome2008}
{Ndome} H.,  {Hochlaf} M.,  {Lewis} B.~R.,  {Heays} A.~N.,  {Gibson} S.~T.,
  {Lefebvre-Brion} H.,  2008, \mn@doi [\jcp] {10.1063/1.2990658}, \href
  {http://adsabs.harvard.edu/abs/2008JChPh.129p4307N} {129, 164307}

\bibitem[\protect\citeauthoryear{Royer}{Royer}{1971}]{royer1971b}
Royer A.,  1971, Phys. Rev. A, 43, 499

\bibitem[\protect\citeauthoryear{Sando \& Wormhoudt}{Sando \&
  Wormhoudt}{1973}]{sando1973}
Sando K.~M.,  Wormhoudt J.~G.,  1973, Phys. Rev. A, 7, 1889

\bibitem[\protect\citeauthoryear{{Shamasundar}, {Knizia}  \&
  {Werner}}{{Shamasundar} et~al.}{2011}]{shamasundar2011}
{Shamasundar} K.~R.,  {Knizia} G.,   {Werner} H.-J.,  2011, \mn@doi [\jcp]
  {10.1063/1.3609809}, \href
  {http://adsabs.harvard.edu/abs/2011JChPh.135e4101S} {135, 054101}

\bibitem[\protect\citeauthoryear{{Spelsberg} \& {Meyer}}{{Spelsberg} \&
  {Meyer}}{2001}]{spelsberg2001}
{Spelsberg} D.,  {Meyer} W.,  2001, \mn@doi [\jcp] {10.1063/1.1400139}, \href
  {http://adsabs.harvard.edu/abs/2001JChPh.115.6438S} {115, 6438}

\bibitem[\protect\citeauthoryear{Spielfiedel}{Spielfiedel}{2003}]{spielfiedel2003}
Spielfiedel A.,  2003, J. Mol. Spectrosc., 217, 162

\bibitem[\protect\citeauthoryear{Spielfiedel, Palmieri  \&
  Mitrushenkov}{Spielfiedel et~al.}{2004}]{spielfiedel2004}
Spielfiedel A.,  Palmieri P.,   Mitrushenkov A.,  2004, Molec. Phys., 102, 2249

\bibitem[\protect\citeauthoryear{{Theodorakopoulos}, {Farantos}, {Buenker}  \&
  {Peyerimhoff}}{{Theodorakopoulos} et~al.}{1984}]{theodora1984}
{Theodorakopoulos} G.,  {Farantos} S.~C.,  {Buenker} R.~J.,   {Peyerimhoff}
  S.~D.,  1984, \mn@doi [Journal of Physics B Atomic Molecular Physics]
  {10.1088/0022-3700/17/8/008}, \href
  {http://adsabs.harvard.edu/abs/1984JPhB...17.1453T} {17, 1453}

\bibitem[\protect\citeauthoryear{Theodorakopoulos, Petsalakis, Nicolaides  \&
  R.J.Buenker}{Theodorakopoulos et~al.}{1987}]{theodora1987}
Theodorakopoulos G.,  Petsalakis I.~D.,  Nicolaides C.~A.,   R.J.Buenker 1987,
  J. Phys. B, 20, 2339

\bibitem[\protect\citeauthoryear{{Tremblay} \& {Bergeron}}{{Tremblay} \&
  {Bergeron}}{2009}]{tremblay2009}
{Tremblay} P.~E.,  {Bergeron} P.,  2009, \mn@doi [Astrophysical Journal]
  {10.1088/0004-637X/696/2/1755}, 696, 1755

\bibitem[\protect\citeauthoryear{{Werner} \& {Knowles}}{{Werner} \&
  {Knowles}}{1985}]{werner1985}
{Werner} H.-J.,  {Knowles} P.~J.,  1985, \mn@doi [\jcp] {10.1063/1.448627},
  \href {http://adsabs.harvard.edu/abs/1985JChPh..82.5053W} {82, 5053}

\bibitem[\protect\citeauthoryear{{Werner} \& {Knowles}}{{Werner} \&
  {Knowles}}{1988}]{werner1988}
{Werner} H.-J.,  {Knowles} P.~J.,  1988, \mn@doi [\jcp] {10.1063/1.455556},
  \href {http://adsabs.harvard.edu/abs/1988JChPh..89.5803W} {89, 5803}

\bibitem[\protect\citeauthoryear{Werner, Knowles, Knizia, Manby, {Sch\"{u}tz}
  et~al.}{Werner et~al.}{2015}]{molpro2015}
Werner H.-J.,  Knowles P.~J.,  Knizia G.,  Manby F.~R.,  {Sch\"{u}tz} M.,
  et~al., 2015, MOLPRO, version 2015.1, a package of ab initio programs

\bibitem[\protect\citeauthoryear{{Xu}, {Zuckerman}, {Dufour}, {Young}, {Klein}
  \& {Jura}}{{Xu} et~al.}{2017}]{xu2017}
{Xu} S.,  {Zuckerman} B.,  {Dufour} P.,  {Young} E.~D.,  {Klein} B.,   {Jura}
  M.,  2017, \mn@doi [\apjl] {10.3847/2041-8213/836/1/L7}, \href
  {http://adsabs.harvard.edu/abs/2017ApJ...836L...7X} {836, L7}

\makeatother
\end{thebibliography}



\section*{appendix}

\begin{figure}
 \centering
\resizebox{0.46\textwidth}{!}
{\includegraphics*{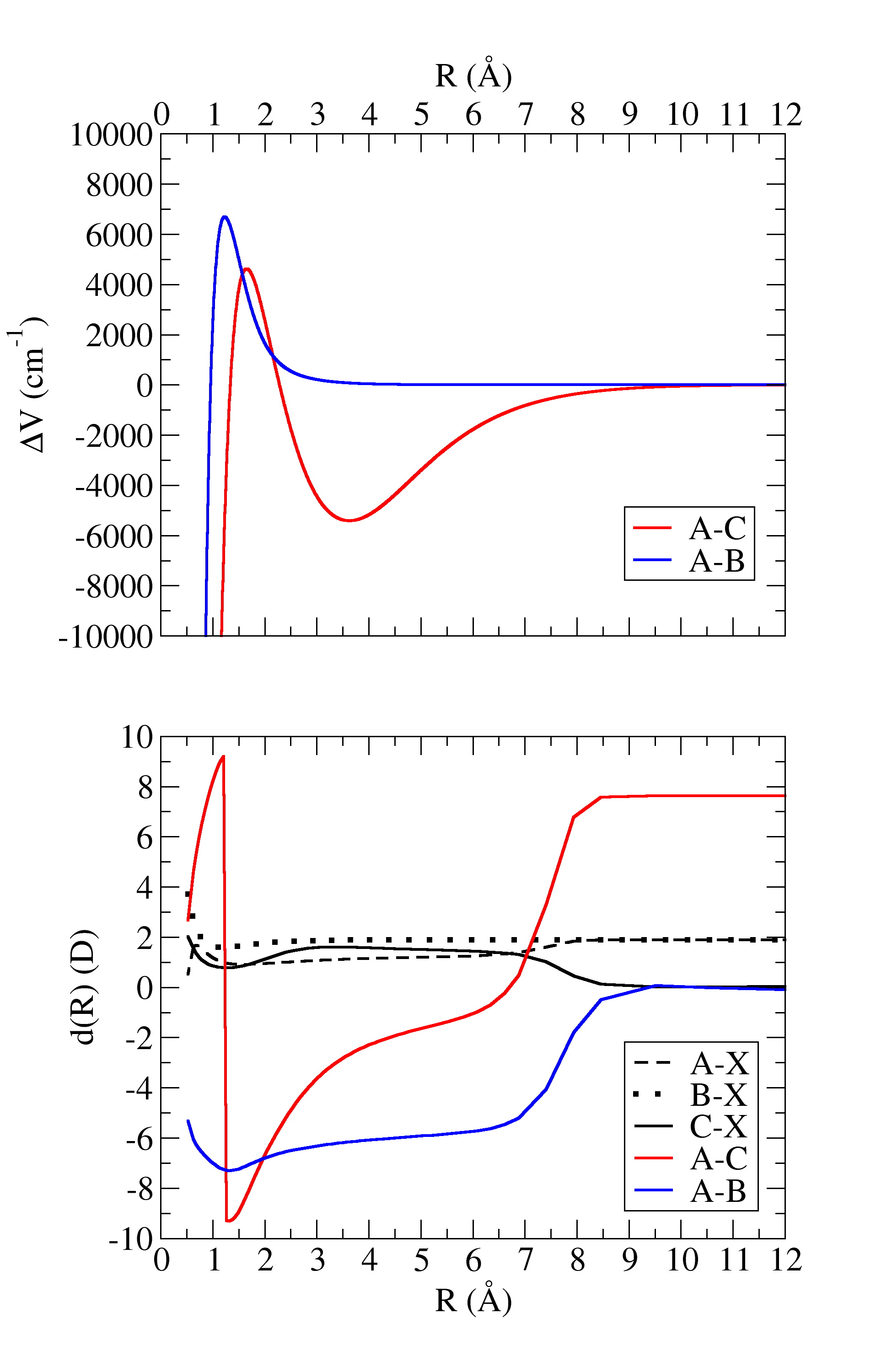}}
\caption  {Potential energy differences in cm$^{-1}$ and electric dipole transition  moments in debye (D or $10^{-18}$ statcoulomb-cm)   between the four lowest electronic states of H--He
  calculated at the MRCI/aug-cc-pV6Z$^\star$ level. Note that the
  \mbox {$C \, \Sigma$ $\leftarrow $ $X \, \Sigma$ } is asymptotically
  forbidden, while transitions between excited states may occur. Upper panel: Energy differences $A \Sigma - B \Sigma$ (blue) and $A\Sigma - C\Pi$ (red).  Lower panel: Electric dipole transition moments for H in the presence of He for states contributing to H~Lyman-$\alpha$. }
\label{dip4}
\end{figure}

Parameters of the H--He molecular potentials are given in Tables~\ref{tab:gaussian} and \ref{tab:energies}. Figure~\ref{dip4} shows the dependence on $R$ of the radiative transition moments between the excited states and the perturbations of those states as the H and He atoms approach from large $R$.

\begin{table}
 \caption{Exponents of the diffuse uncontracted Gaussian
primitives added to the aug-cc-pV6Z basis set to form the presently used
aug-cc-pV6Z* basis sets for the H and He atoms.}
 \label{tab:gaussian}
 \begin{tabular}{lccc}
  \hline
  State & 1 & 2 & 3\\
H(\emph{s}) & 0.00690204 & 0.002520537 & 0.000920468\\
H(\emph{p}) & 0.026565598 & 0.010533298 & 0.004176468\\
H(\emph{d}) & 0.055406537 & 0.024364162 & 0.010713761\\
H(\emph{f}) & 0.106396067 & 0.046204584 & 0.020065249\\
H(\emph{g}) & 0.168703345 & 0.069928301 & 0.028985598\\
H(\emph{h}) & 0.175320015 & 0.045069073 & 0.011585793\\
He(\emph{s}) & 0.017177900 & 0.006596920 & 0.002533450\\
He(\emph{p}) & 0.050416903 & 0.019858313 & 0.007821833\\
He(\emph{d}) & 0.094209988 & 0.036827891 & 0.014396494\\
He(\emph{f}) & 0.151890237 & 0.056684629 & 0.021154402\\
He(\emph{g}) & 0.232902520 & 0.079072280 &  0.026845675\\
He(\emph{h}) & 0.248198125 & 0.060632194 & 0.014811808\\
  \hline
 \end{tabular}
\end{table}

\begin{table}
 \caption{Dissociation fragments, experimental and calculated relative dissociation asymptotic energies, and molecular states
for
the four lowest electronic states of H--He.
Experimental data are from \citet{kramida2010}.
}

\label{tab:energies}
 \begin{tabular}{ccccc}
  \hline
 Atomic &  & Observed & Calculated  & Molecular \\
 H & He & (cm$^-1$) & (cm$^-1$) &\\
 \hline
$1s\,^2S_g$   &  $1s^2\,^1S_g$  & 0$^a$          & 0$^a$     & $X\,^2\Sigma^+$\\
$2p\,^2P_u$   &  $1s^2\,^1S_g$  & 82259          & 82308     & $A\,^2\Sigma^+$, $B \, ^2\Pi$\\
$2s\,^2S_g$   &  $1s^2\,^1S_g$  & 82259          & 82308     & $C\,^2\Sigma^+$\\
  \hline
$^a$Reference \\  
 \end{tabular}
\end{table}

\begin{table}
 \caption{Spectroscopic constants and dissociation energies for
the three lowest excited electronic states of H--He as deduced from the
MRCI+Q /aug-cc-pV6Z* potential energy curves. $R_e$
corresponds to the equilibrium distance. $\omega_e$ and
$\omega_e x_e$ are the vibrational constants.
$\beta_e$ and $\alpha_e$ are the rotational constants.
$D_e$ is the dissociation energy.}
\label{tab:constants}
\begin{tabular}{ccccccc}
\hline
State & $R_e$  & $\omega_e$  & $\omega_e x_e$  & $\beta_e$ & $\alpha_e$ & $D_e$ \\
& \AA & cm$^{-1}$ &cm$^{-1}$ &cm$^{-1}$  &cm$^{-1}$ & eV \\
\hline
$A\,^2\Sigma^+$ & 0.74074 & 3697.2 & 149.5 & 38.16 & 2.608 & 2.563\\
$B\,^2\Pi$ & 0.76863 & 3313.4 & 149.8 & 35.44 & 2.629 & 2.218\\
$C\,^2\Sigma^+$& 0.80953 & 2906.3 & 144.0 & 31.95 & 2.551 & 1.638\\
\hline
 \end{tabular}
\end{table}

\bsp	
\label{lastpage}
\end{document}